\documentstyle[titlepage,preprint,aps,floats,epsfig]{revtex}
\newcommand{\beq}{\begin{equation}}
\newcommand{\eeq}{\end{equation}}
\newcommand{\eq}[1]{Eq.(\ref{#1})}

\begin{document}

\draft
\tighten
\preprint{UK/01-01}
\title {One-Loop Electron Vertex in Yennie Gauge}
\medskip
\author {Michael I. Eides \thanks{E-mail
address:  eides@pa.uky.edu, eides@thd.pnpi.spb.ru}}
\address{Department of Physics and Astronomy, University of Kentucky,
Lexington, KY 40506, USA \\
and Petersburg Nuclear Physics Institute, Gatchina,
St.Petersburg 188350, Russia}
\author{Valery A. Shelyuto \thanks{E-mail address: shelyuto@vniim.ru}}
\address{D. I. Mendeleev Institute of
Metrology, St.Petersburg 198005, Russia}

\date{February, 2001}

\maketitle

\begin{abstract}
We derive a compact Yennie gauge representation for the off-shell
one-loop electron-photon vertex, and discuss it properties. This
expression is explicitly infrared finite, and it has proved to be
extremely useful in multiloop calculations in the QED bound state
problem.
\end{abstract}

\section{Introduction}

As is well known the physical results in quantum electrodynamics (as in
any gauge theory) are gauge invariant, but the calculations themselves
are gauge dependent. A proper choice of gauge may greatly facilitate
calculations of radiative corrections in a specific problem.

In quantum electrodynamics gauge freedom is described by
the transformation of the photon propagator

\beq
D_{\alpha \beta} (q)\,\rightarrow\, D_{\alpha \beta} (q)
 \,+\, q_{\alpha}{\chi}_{\beta} \,+\, {\chi}_{\alpha} q_{\beta}~ ,
\eeq

\noindent
where $~{\chi}_{\alpha}~$ are arbitrary functions of momentum $~q~$.

While the full gauge invariant sets of diagrams which describe the
physical processes are gauge independent, the individual diagrams and
the complexity of calculations strongly depend on the choice of gauge.
The infrared safe Yennie gauge \cite{abr,friedyennie} defined by the
photon propagator

\beq
D_{\alpha \beta} (q) = \frac{1}{q^2 + i\varepsilon} \,
\left(g_{\alpha \beta} \, +  \,
\frac{2 q_{\alpha} q_{\beta} }{q^2} \right) ~~
\eeq

\noindent
is particularly well suited for the bound state problems, where it
greatly alleviates the notorious infrared difficulties specific for
such kind of problems (see, e.g.,
\cite{tomozawa,sty,eksyennie,eksann1,eksann2}).

There is no infrared photon radiation in the bound state problems, and
all infrared divergences should cancel in the final results. The most
useful technically feature of the Yennie gauge, which is shares with
the noncovariant Coulomb gauge, is that the infrared behavior of the
individual diagrams is greatly improved in comparison with the infrared
behavior of the diagrams in other covariant gauges. In
particular many diagrams, which are infrared divergent in other
relativistic gauges (Feynman, Landau, etc.), are infrared finite in the
Yennie gauge. This feature of the Yennie gauge allows to perform
explicitly covariant calculations without introducing an intermediate
infrared photon mass, which is inevitable in other common relativistic
gauges. Thus the Yennie gauge combines the nice infrared properties of
the noncovariant Coulomb gauge (see, e.g., \cite{adkins83,adkins86})
with the advantages specific to the explicitly Lorentz covariant
gauges.

The Yennie gauge is widely used in the bound state theory (see, e.g.,
\cite{tomozawa,sty,eksyennie,eksann1,eksann2}, and references in
\cite{review}). In the framework of dimensional regularization one- and
two-loop calculations in the Yennie gauge were discussed in
\cite{adkins1,adkins2,adkins3,adkins4}. The Yennie gauge was
extensively used in our papers on the one- and two-loop
radiative corrections to the bound state energy levels
\cite{eksann1,eksann2,beks,esjetp,es,egs00}. We have obtained a compact
infrared soft integral representation for the renormalized one-loop
vertex in the Yennie gauge, which  turned out to be extremely useful in
calculations. Below we will derive this representation for the
Yennie gauge vertex, and discuss its main features.

\section{Infrared Finite Bare Vertex}

\begin{figure}[ht]
\centerline{\epsfig{file=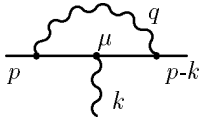}}
\vspace{0.5cm}
\caption{One-loop vertex}
\label{vert}
\end{figure}

General expression for the off-mass-shell one-loop vertex in the Yennie
gauge (see Fig.~\ \ref{vert}) has the form

\beq  \label{genexp}
\Lambda_{\mu} (p,p-k) = \frac{\alpha}{4 \pi} \int \frac{d^4 q}{\pi^2 i}
\, \frac{\gamma^{\alpha} (\hat{p} + \hat{q} + m) \gamma_{\mu}
(\hat{p} + \hat{q} - \hat{k} + m) \gamma^{\beta}}
{ D(p+q) D(p+q-k)  } \,  D_{\alpha \beta} (q)~,
\eeq

\noindent
where

\beq
D(p) = p^2 - m^2 + i\varepsilon ~.
\eeq

We would like to obtain an integral representation for the Yennie
gauge vertex which is explicitly infrared finite. Ultraviolet
divergences in the Yennie gauge should be subtracted as usual, and to
this end it is convenient to have a simple expression for the
ultraviolet divergent term.  Let us first separate the ultraviolet
divergent contributions. They are generated by the large integration
momenta $q \to \infty$ in \eq{genexp} when the Feynman and
longitudinal parts of the numerator may be written as

\[
\gamma^{\alpha} (\hat{p} + \hat{q} + m) \gamma_{\mu}
(\hat{p} + \hat{q} - \hat{k} + m) \gamma_{\alpha}
~\simeq ~ q^2 \gamma_{\mu}
~\simeq ~ D(p+q-k) \gamma_{\mu}~,
\]
\beq           \label{uvterm}
\frac{2}{q^2}~ \hat{q} (\hat{p} + \hat{q} + m) \gamma_{\mu}
(\hat{p} + \hat{q} - \hat{k} + m) \hat{q}
~\simeq ~ 2 q^2 \gamma_{\mu}
~\simeq ~ 2D(p+q-k) \gamma_{\mu}~.
\eeq

\noindent
We have represented the large terms in the numerator  as coefficients
before the electron denominator $D(p+q-k)$ in order to get rid of the
$k$ dependence in the ultraviolet divergent contributions.

The Yennie gauge vertex is infrared finite, and we are looking for such
representation of the electron-photon vertex where all would be
infrared divergences cancel already in the integrand. The most
infrared singular terms in the integrand in \eq{genexp} correspond to
the terms in the numerator which do not contain the integration
momentum $q$

\beq           \label{irterm}
\gamma^{\alpha} (\hat{p} + \hat{q} + m) \gamma_{\mu}
(\hat{p} + \hat{q} - \hat{k} + m) \gamma^{\beta}
~\simeq ~ \gamma^{\alpha} (\hat{p} + m) \gamma_{\mu}
(\hat{p} - \hat{k} + m) \gamma^{\beta}~.
\eeq

\noindent
Separating the ultraviolet and infrared divergent contributions
we write the integral representation for the Yennie gauge vertex in
\eq{genexp} in the form

\beq  \label{speuvir}
\Lambda_{\mu} (p,p-k) = \frac{\alpha}{4 \pi} \int \frac{d^4 q}{\pi^2 i}
\biggl\{~ 3 \gamma_{\mu} H(0,1,1) ~+~ N_1 H(1,1,1)
\eeq
\[
+~~ \gamma^{\alpha} (\hat{p} + m) \gamma_{\mu}
(\hat{p} - \hat{k} + m) \gamma^{\beta} \biggl[~ g_{\alpha \beta} H(1,1,1)
~+~ 2 q_{\alpha} q_{\beta} H(1,1,2)~\biggr]~\biggr\}~,
\]

\noindent
where

\beq
N_1 ~=~ 3q^2 \gamma_{\mu}
~-~ 3 D(q+p-k) \gamma_{\mu}
~+~\gamma_{\alpha} \hat{q} \gamma_{\mu}
(\hat{p} - \hat{k} + m) \gamma^{\alpha}
~+~\gamma_{\alpha} (\hat{p} + m) \gamma_{\mu} \hat{q} \gamma^{\alpha}
\eeq
\[
~+~2\gamma_{\mu}( \hat{p}  - \hat{k} + m) \hat{q}
~+~2\hat{q}(\hat{p} + m) \gamma_{\mu}~
~-~ (2 \hat{q} \gamma_{\mu} \hat{q} + q^2 \gamma_{\mu})~,
\]

\noindent
and

\beq
H(m,1,n) \equiv  \bigl\{D^m(q+p-k)\, D(q+p) \,
q^{2n} \bigr\}^{-1} ~.
\eeq

\noindent
All ultraviolet  divergent contributions in \eq{speuvir} correspond to
the numerators in \eq{uvterm}, and are collected in the term with the
denominator $H(0,1,1)$, which is independent of the transferred
momentum $k$, and depends only on the external fermion momentum $p$.
All potentially infrared divergent contributions to the integral
generated by the Feynman and the longitudinal terms in the virtual
photon propagator correspond to the numerator in \eq{irterm}, and are
collected in the last term with the square brackets in the integrand in
\eq{speuvir}.  All other terms in this integrand are explicitly
infrared finite since they either contain an extra power of the virtual
momentum $q$ in the numerator, or respective denominators are less
singular at small $q$.

The final integral representation will be written in terms of the
combinations of external momenta which naturally arise when we combine
the Feynman denominators. For example,

\beq
H(1,1,1) = \int^1_0 {dx} \int^1_0 {dz} ~
\frac {2x}{\bigl[(q-xQ)^2 - x\Delta \bigr]^3}~,
\eeq

\noindent
where

\beq
Q = -p + kz~,
\eeq

\beq
\Delta = m^2 - k^2 z(1-xz) + 2pk (1-x)z - p^2 (1-x)~.
\eeq

\noindent
The ultraviolet divergent contribution generated by the term with
the denominator $H(0,1,1)$ will be written in terms of the degenerate
function $\Delta_0$

\beq
\Delta_0 \equiv \Delta (z=0) = m^2 x - D(p)(1-x)~.
\eeq

\noindent
After the shift of the integration variable $q ~\to ~ q + xQ$ we obtain

\[
\Lambda_{\mu} (p,p-k) = \frac{\alpha}{4 \pi} \int \frac{d^4 q}{\pi^2 i}
\biggl\{3 \gamma_{\mu} \bar{H}(0,1,1)
+ \bar{N}_1 \bar{H}(1,1,1) + 2 x^2 N_2  \bar{H}(1,1,2)
\]
\[
+~~ \gamma_{\alpha} (\hat{p} + m) \gamma_{\mu}
(\hat{p}-\hat{k}+m) \gamma^{\alpha}
\biggl[ \bar{H}(1,1,1)
~+~ \biggl(\frac{q^2}{2} + 2x^2 Q^2 \biggr) \bar{H}(1,1,2)
\biggr]~\biggr\}~~
\]
\[
= \frac{\alpha}{4 \pi}
\int^1_0 {dx} \int^1_0 {dz}
\biggl\{~ 3 \gamma_{\mu} \cdot
\biggl[~\ln{\frac{\Lambda^2}{x \Delta_0}} ~-~1~\biggr]
~-~ \frac{\bar{N}_1}{\Delta}  ~+~ 2N_2 \frac{x(1-x)}{\Delta^2}
\]
\beq              \label{infrcanc}
+~~ \gamma_{\alpha} (\hat{p} + m) \gamma_{\mu}
(\hat{p} - \hat{k} + m) \gamma^{\alpha}
\biggl[-\frac{1}{\Delta} - \frac{1-x}{\Delta}
+ \frac{2x(1-x)Q^2}{\Delta^2} \biggr]~\biggr\}~,
\eeq

\noindent
where $~\Lambda~$ is the ultraviolet cutoff,

\[
\bar{N}_1=-~ 3 D(p-k) \gamma_{\mu}
~+~ x \biggl[-6(p-k)Q \gamma_{\mu}
+ \gamma_{\alpha}\hat{Q}\gamma_{\mu} (\hat{p} - \hat{k} + m)
\gamma^{\alpha}
\]
\beq
+ \gamma_{\alpha} (\hat{p} + m)\gamma_{\mu}\hat{Q} \gamma^{\alpha}
+~ 2 \gamma_{\mu} (\hat{p} - \hat{k} + m) \hat{Q}
+ 2 \hat{Q} (\hat{p} + m)\gamma_{\mu}\biggr]
~-~ x^2 \biggl(2 \hat{Q} \gamma_{\mu} \hat{Q} + Q^2 \gamma_{\mu}
\biggr)~,
\eeq
\beq
N_2 ~=~ \hat{Q} (\hat{p} + m) \gamma_{\mu} (\hat{p} - \hat{k} + m) \hat{Q}
~-~ Q^2 \gamma_{\alpha} (\hat{p} + m) \gamma_{\mu}
(\hat{p} - \hat{k} + m) \gamma^{\alpha}~,
\eeq

\noindent
and function $\bar{H}(m,1,n)$ is just the  function ${H}(m,1,n)$
after the shift of the integration momentum $q ~\to ~ q + xQ$.

The term with the numerator $N_2$ in \eq{infrcanc} is the price we have
to pay for the simple $\gamma$-matrix structure of the last term in the
square brackets in \eq{infrcanc}. The point is that a more complicated
matrix structure of the form $\hat{Q}(\dots)\hat{Q}$ arises naturally
after the shift $q ~\to ~ q + xQ$. However, the trivial structure
proportional to $Q^2$ greatly facilitates consideration of the would
be infrared divergences, and we simply wrote infrared singular terms in
the square brackets in \eq{infrcanc} in a convenient form, and collected
the compensating infrared safe terms in the numerator $N_2$.

The contributions generated by the terms $1/\Delta$ and $x/\Delta^2$ in
the last square brackets in \eq{infrcanc} are infrared divergent on
the mass shell at zero momentum transfer. Really, these terms behave
as $1/x$ if $p^2 \to m^2$ and $k=0$, and thus are infrared
divergent. In the Feynman gauge the vertex itself is also infrared
divergent under these conditions but in the Yennie gauge infrared
divergent contributions corresponding to the Feynman and longitudinal
parts of the photon propagator cancel each other. In order to
cancel these apparent infrared divergences we use the identity

\beq  \label{ident}
\int^1_0 {dx} ~\frac{\partial}{\partial x}
\biggl\{\frac{x(1-x)}{\Delta}\biggr\} =
\int^1_0 {dx} ~ \biggl\{\frac{1-2x}{\Delta} -
\frac{x(1-x)Q^2}{\Delta^2}\biggr\} = 0 ~,
\eeq

\noindent
which may be easily proved with the help of the relation  $\partial
\Delta/\partial x ~=~ Q^2$. Applying this identity we see that the
infrared divergent contributions corresponding to the Feynman and
longitudinal terms in the photon propagator cancel each other $-1 -
(1-x) + 2(1-2x) = - 3x$. Then the sum of the would be divergent
contributions is reduced to the infrared safe form $-3xm^2/\Delta$,
which is finite on the mass shell at zero momentum transfer.

After these transformations the Yennie gauge electron-photon vertex
may be written in a compact form

\[
\Lambda_{\mu} (p,p-k) = \frac{\alpha}{4\pi}
\int^1_0 {dx} \int^1_0 {dz}\biggl\{~ 3 \gamma_{\mu}
\biggl[~\ln{\frac{\Lambda^2}{m^2x^2}} ~-~ 1
~+~ \frac{D(p)}{\Delta_0} ~\biggr]
\]
\beq                 \label{finalbare}
-~ \frac{\bar{N}_1}{\Delta}
~-~ 3 \gamma_{\alpha} (\hat{p} + m) \gamma_{\mu}
(\hat{p} - \hat{k} + m) \gamma^{\alpha} \frac{x}{\Delta}
~+~ 2 N_2 \frac{x(1-x)}{\Delta^2}~\biggr\}~.
\eeq

\section{Renormalization}

As usual in QED to obtain the renormalized one-loop vertex from the
expression in \eq{finalbare} we use mass shell subtraction at zero
momentum transfer. However, due to absence of the infrared
regularization the infrared finiteness of the subtraction term is not
guaranteed, and we should first check that it is infrared finite.
Consider asymptotic behavior of the vertex in \eq{finalbare} at $\rho_1
\equiv1-p^2/m^2  \to 0$, $~\rho_2 \equiv 1-(p-k)^2/m^2 \to 0$ and small
but nonzero momentum transfer squared $k^2$. The largest contributions
to the vertex in this regime have the form $k^2\ln{\rho_i}$, and they
are generated by the term with $N_2/\Delta^2$ in the integrand in
\eq{finalbare}. Throwing away all contributions which are at least
linear in the virtualities of the electron lines, i.e., terms of the
form $\rho_1$, $\rho_2$, $(\hat p - m) \simeq -m \rho_1 /2$, and
$(\hat p - k - m) \simeq -m \rho_2 /2$, we obtain

\beq      \label{n2ir}
N_2 \simeq  4 \gamma_{\mu} \Bigl[pk(-p^2+pkz)
+(pk)^2z(-3+z)+p^2 k^2 z(1-z)\Bigr]~.
\eeq

\noindent
It is easy to see that $k^2-2pk \to 0$ at $\rho_1\to 0$ and $\rho_2\to
0$. Then we substitute $2pk\to k^2$ in \eq{n2ir}, and preserving only
the leading in $k^2$ terms we have

\beq
N_2 \simeq 2[1+2z(1-z)]m^2k^2\gamma_{\mu}
\to \frac{8}{3}~ m^2 k^2 \gamma_{\mu}~,
\eeq

\noindent
where we have effectively integrated over
$z$ on the right hand side. Integrating also over $x$ we obtain

\beq
\frac{\alpha}{4 \pi} \int^1_0 {dx} \int^1_0 {dz} ~
\frac{2x(1-x)N_2}{\Delta^2} ~\simeq ~
\frac{4\alpha}{3\pi} \frac{k^2}{m^2} \gamma_{\mu} \int^1_0 {dz}
\ln{\frac{1}{(1-z)\rho_1 +z\rho_2}}
\eeq
\[
~\simeq ~ \frac{4\alpha}{3\pi} \frac{k^2}{m^2} \gamma_{\mu}
\ln{\frac{1}{max(\rho_1, \rho_2)}}~.
\]

\noindent
This asymptotic behavior demonstrates that the Yennie gauge vertex
admits subtraction on the mass shell without any additional infrared
regularization. The only subtlety is that we first should put the
momentum transfer squared to be zero, and only then go on the mass
shell.

Let us calculate the subtraction constant. The numerator structures
in \eq{finalbare} simplify at $~k = 0~$ and  $~ \hat{p} = m~$:

\beq
-~\bar{N}_1 ~ \to ~ 3x(2+x) m^2 \gamma_{\mu}~,
\eeq
\beq
-~3 x \gamma_{\alpha} (\hat{p} + m) \gamma_{\mu}(\hat{p} + m)
\gamma^{\alpha}
~ \to ~ -~12 x m^2 \gamma_{\mu}~,
\eeq
\beq
N_2 ~\to ~ 0~~,~~~~~~ \Delta ~\to~ \Delta_0  ~\to~ m^2 x ~.
\eeq

\noindent
Then the infrared finite subtraction constant may be easily
calculated

\[
\Lambda_{\mu} (m,m) = \gamma_{\mu} \frac{3\alpha}{4\pi} \int^1_0 {dx}~
\biggl\{~\ln{\frac{\Lambda^2}{m^2x^2}} ~-~ 1
~+~ \frac{3x(2+x)}{x} ~-~ \frac{12x}{x}~\biggr\}
\]
\beq
= ~\gamma_{\mu}\frac{3\alpha}{4\pi}
\biggl(~\ln{\frac{\Lambda^2}{m^2}} ~-~ \frac12~\biggr)
~\equiv ~ \gamma_{\mu} \bigl( - 1 + Z^{-1}_{1} \bigr)~.
\eeq

\noindent
The final expression for the unrenormalized Yennie gauge vertex is

\beq
\Lambda_{\mu}(p,p-k) = \gamma_{\mu} \bigl( - 1 + Z^{-1}_{1} \bigr)
~+~ \Lambda_{\mu}^R (p,p-k)~.
\eeq

\noindent
The ultraviolet and infrared finite renormalized electron-photon
vertex in the Yennie gauge has the form

\beq               \label{renormvert}
\Lambda_{\mu}^R (p,p-k) = \frac{\alpha}{4 \pi}
\int^1_0 {dx} \int^1_0 {dz}
\biggl\{~ \frac{F^{(0)}_{\mu}}{\Delta_0}
~+~ \frac{F^{(1)}_{\mu}}{\Delta}
~+~ \frac{F^{(2)}_{\mu}}{\Delta^2}~\biggr\}~,
\eeq

\noindent
where

\beq
F^{(0)}_{\mu} = 3\gamma_{\mu}D(p)~,
\eeq
\[
F^{(1)}_{\mu} = 3\gamma_{\mu}\Bigr[D(p-k)
+(2-x)\Delta \Bigr]
- x\Bigl[3 \gamma_{\alpha}(\hat{p}+m)\gamma_{\mu}
(\hat{p}-\hat{k}+m)\gamma^{\alpha}
\]
\[
-6(p-k)Q \gamma_{\mu}
+\gamma_{\alpha}\hat{Q}\gamma_{\mu}(\hat{p}-\hat{k}+m)\gamma^{\alpha}
+\gamma_{\alpha}(\hat{p}+m)\gamma_{\mu}\hat{Q}\gamma^{\alpha}
\]
\beq
+ 2\gamma_{\mu}(\hat{p}-\hat{k}+m)\hat{Q}
+2\hat{Q}(\hat{p}+m)\gamma_{\mu}\Bigr]
+x^2 \Bigl(2\hat{Q}\gamma_{\mu}\hat{Q}
+ Q^2 \gamma_{\mu} \Bigr)~,
\eeq
\beq
F^{(2)}_{\mu} = 2x(1-x)\Bigl[\hat{Q} (\hat{p} + m) \gamma_{\mu} (\hat{p} -
\hat{k} + m) \hat{Q} ~-~ Q^2 \gamma_{\alpha} (\hat{p} + m) \gamma_{\mu}
(\hat{p} - \hat{k} + m) \gamma^{\alpha}\Bigr]~.
\eeq

It is not difficult to check cancellation of the infrared finite
renormalization constants $Z_1$ and $Z_2$ in the Yennie gauge.
The explicit expression for the one-loop self-energy operator in the
Yennie gauge is well known (see, e.g., \cite{tomozawa,eksann1})

\beq
\Sigma (p) ~=~ \delta m
~+~ \bigl( - 1 + Z^{-1}_{2} \bigr) (\hat{p} - m) ~+~ \Sigma^R (p)~,
\eeq

\noindent
where the renormalized self-energy operator has the form

\beq     \label{selfenren}
\Sigma^R (p) ~=~ \frac{\alpha}{4 \pi} (\hat{p}-m)^2
\int_0^1 {dx}~\frac{- 3 \, \hat{p}\,x}{m^2 x - D(p)(1-x)}~,
\eeq

\noindent
the mass renormalization is given by the expression

\beq
\delta m ~=~ \frac{3\alpha}{4\pi}
\biggl(~\ln{\frac{\Lambda^2}{m^2}} ~+~ \frac12~\biggr)~,
\eeq

\noindent
and the wave function renormalization constant has the form

\beq
 1 - Z^{-1}_{2} ~\equiv ~  \Sigma^{'} (m) ~=~
-~\frac{3 \alpha}{4 \pi}
\biggl(~\ln{\frac{\Lambda^2}{m^2}} ~-~ \frac12~\biggr)~.
\eeq

\noindent
It is easy to see that the Ward identity is satisfied, and the
infrared finite renormalization constants $Z_1$ and $Z_2$ coincide

\beq
\Lambda_{\mu}(m,m)~=~-~\Sigma^{'} (m)~,~~~~~~~~
Z_1 ~=~ Z_2~.
\eeq

\section{Infrared and Ultraviolet Asymptotic Behavior of the Yennie
Gauge Vertex}

The integral representation for the Yennie gauge vertex in
\eq{renormvert} is most convenient for calculations of radiative
corrections, and the usual representation of the vertex in terms of the
Lorentz invariant form factors is neither necessary nor calculationally
useful. However, quite often in the bound state problems one needs to
treat separately the terms in the vertex which have different
asymptotic behavior at small momentum transfer (see, e.g.,
\cite{eksyennie,eksann1}). All terms in the Yennie gauge vertex besides
the anomalous magnetic moment contribution vanish at least as momentum
transfer squared when $k^2\to0$. The anomalous magnetic moment
contribution is linear in the momentum transfer, and it determines the
asymptotic behavior of the Yennie gauge vertex at small momentum
transfer. It is not difficult to identify the anomalous magnetic moment
contribution in \eq{renormvert}. All terms which contribute to the
anomalous magnetic moment may be extracted from the term with the
numerator $-\bar{N}_1$ in \eq{infrcanc}

\beq
-x \gamma_{\alpha}\hat{Q}\gamma_{\mu} (\hat{p} - \hat{k} + m)
\gamma^{\alpha} ~\to ~ -2x(1-z) m \sigma_{\mu \nu} k^{\nu}~,
\eeq
\beq
-x \gamma_{\alpha} (\hat{p} + m)\gamma_{\mu}\hat{Q} \gamma^{\alpha}
~ \to ~ -2xz m \sigma_{\mu \nu} k^{\nu}~,
\eeq
\beq
2x^2 \hat{Q} \gamma_{\mu} \hat{Q}
~\to ~ 2x^2 m \sigma_{\mu \nu} k^{\nu}~.
\eeq

\noindent
At small momentum transfer ($k\to0$) on the mass shell ($p^2=m^2$)
the denominator $\Delta \to m^2 x$, and the sum of these terms
generates the anomalous magnetic moment

\beq
-~\frac{\alpha}{2 \pi} \frac{\sigma_{\mu \nu} k^{\nu}}{2m}~.
\eeq

\noindent
Separating the anomalous magnetic moment we write the renormalized
one-loop vertex in the form

\beq
\Lambda_{\mu}^R(p,p-k) ~=~{\tilde \Lambda}_{\mu}^R (p,p-k)
~-~\frac{\alpha}{2 \pi} \frac{\sigma_{\mu \nu} k^{\nu}}{2m}~.
\eeq

\noindent
The scalar factors before the tensor structures in ${\tilde
\Lambda}_{\mu}^R (p,p-k)$ depend only on the momentum transfer squared
$k^2$ and the  electron line virtuality $\rho = (m^2-p^2)/m^2$. At
small momenta transfer and near the mass shell all entries in the
expression for ${\tilde \Lambda}_{\mu}^R (p,p-k)$ are either linear in
$k^2$ and/or $\rho$, or are proportional to the projector $\hat
p-m$ on the mass shell. At $k=0$ we have

\beq
{\tilde \Lambda}_{\mu}^R (p,p)
~\simeq ~-\gamma_{\mu} \frac{3\alpha}{4\pi}~ \rho
~\simeq~  \gamma_{\mu} \frac{3\alpha}{2\pi}~\frac{\hat{p}-m}{m}~.
\eeq

\noindent
According to the Ward identity

\beq
{\tilde \Lambda}_{\mu}^R (p,p)
~=~ - \frac{\partial \Sigma^R (p)}{\partial p^{\mu}}~,
\eeq

\noindent
the small  momentum transfer behavior of the vertex is connected
with the behavior of the self-energy operator near the mass
shell. The mass operator in \eq{selfenren} near the mass shell $\hat{p}
\to m$ is

\beq
\Sigma^R (p) ~\simeq ~ -\frac{3\alpha}{4\pi}~\frac{(\hat{p}-m)^2}{m}~,
\eeq

\noindent
and it is easy to see that the Ward identity near the mass shell is
satisfied.

We had used the Yennie gauge one-loop electron-photon vertex
$\Lambda_{\mu}^R (p+q,p+q-k)$ from \eq{renormvert} as a subdiagram in
two-loop calculations \cite{esjetp,es}. In such calculations not only
the infrared but also the ultraviolet behavior of the one-loop vertex
at $-q^2 \to \infty$ should be under control. The dominant logarithmic
contribution to $\Lambda_{\mu}^R (q,q)$ is generated in this regime
exclusively by the first term in the braces in \eq{renormvert}

\beq     \label{uvasympt}
\Lambda_{\mu}^R (q,q) \simeq - \frac{3\alpha}{4\pi}\gamma_{\mu}
\ln{\frac{-q^2}{m^2}}~.
\eeq

\noindent
All other contributions are nonlogarithmic. Ultraviolet behavior of the
Yennie gauge vertex is no better than the ultraviolet behavior in the
Feynman gauge, and really the contribution in \eq{uvasympt} differs
from the respective Feynman gauge expression only by a multiplicative
factor $3$. As is well known the ultraviolet vertex logarithm
$\ln{(-q^2/m^2)}$ does not arise in the Landau gauge, which is the most
convenient gauge for extracting the large ultraviolet logarithms.

Let us consider in more detail behavior of the different entries in the
integrand in \eq{renormvert} at $-q^2 \to \infty$

\beq
F^{(1)}_{\mu} \simeq -3(1-x)^2 q^2 \gamma_{\mu}
+ x(1+x)\Bigl[q^2 \gamma_{\mu} + 2\hat{q}\gamma_{\mu}\hat{q} \Bigr]~ ,
\eeq
\beq
F^{(2)}_{\mu} \simeq 2x(1-x) q^2
\Bigl[q^2 \gamma_{\mu} + 2\hat{q}\gamma_{\mu}\hat{q} \Bigr] ~,
\eeq
\beq
\Delta \simeq -(1-x)q^2 ~.
\eeq

\noindent
It is easy to see that in this regime the finite integral

\beq   \label{uvintex}
\int^1_0 {dx} \biggl\{~ \frac{F^{(1)}_{\mu}}{\Delta}
~+~ \frac{F^{(2)}_{\mu}}{\Delta^2}~\biggr\} \simeq
\int^1_0 {dx} \biggl\{~ \frac{2}{-(1-x)q^2}
~+~ \frac{2(1-x)q^2}{(1-x)^2 q^4}~\biggr\}
\Bigl[q^2 \gamma_{\mu} + 2\hat{q}\gamma_{\mu}\hat{q} \Bigr]
\eeq

\noindent
is a sum of two integrals over the Feynman parameter $x$ each of which
diverges at $x \to 1$. In calculations of the two-loop radiative
corrections we need to integrate this expression over the momentum $q$.
Then it is often convenient and even necessary to consider the two
terms in \eq{uvintex} separately. Note first that the factor

\beq
\Bigl[q^2 \gamma_{\mu} + 2\hat{q}\gamma_{\mu}\hat{q} \Bigr] ~\simeq ~
\biggl[q^2 \gamma_{\mu} + \frac{2(-2)}{4} q^2 \gamma_{\mu}\biggr]
\eeq

\noindent
vanishes after integration over $q$ in the two-loop diagrams, if
all other factors in the integrand depend only on $q^2$, but this is
often not the case. Then one needs to avoid the spurious divergences at
$x \to 1$ by rearranging different terms in \eq{renormvert} with the
help of the identity in \eq{ident}, which we already used to improve
the infrared behavior. After transformations we obtain a slightly
different representation  for the renormalized vertex in the Yennie
gauge

\beq     \label{vertuvimp}
\Lambda_{\mu}^R (p,p-k) = \frac{\alpha}{4 \pi}
\int^1_0 {dx} \int^1_0 {dz}
\biggl\{~ \frac{F^{(0)}_{\mu}}{\Delta_0}
~+~ \frac{{\tilde F^{(1)}_{\mu}}}{\Delta}
~+~ \frac{{\tilde F^{(2)}_{\mu}}}{\Delta^2}~\biggr\}~,
\eeq

\noindent
where

\beq \label{transfff}
{\tilde F^{(1)}_{\mu}} =  F^{(1)}_{\mu} +
2(1-2x)\Bigl[q^2 \gamma_{\mu} + 2\hat{q}\gamma_{\mu}\hat{q} \Bigr]~ ,
\eeq
\beq
{\tilde F^{(2)}_{\mu}} =  F^{(2)}_{\mu} -
2x(1-x)Q^2\Bigl[q^2 \gamma_{\mu} + 2\hat{q}\gamma_{\mu}\hat{q} \Bigr]~ .
\eeq

\noindent
The transformation of the numerator structures in \eq{transfff} does
not change the infrared behavior of the vertex. Which of the
representations \eq{renormvert} and  \eq{vertuvimp} to use in
calculations of the two-loop corrections depends on the nature of the
two-loop diagram. For example, in calculation of the contributions of
order $\alpha^2(Z\alpha)^5$ generated by the diagrams with the vertex
insertions in the ultraviolet divergent skeleton diagrams (see
e.g., diagrams $(i,m,o)$ in \cite{esjetp,es}) the representation in
\eq{vertuvimp} is more convenient.  On the other hand, the
representation in \eq{renormvert} is more convenient for calculation of
the contributions generated by the diagrams with the vertex insertions
in the ultraviolet finite skeleton diagrams (see, e.g., diagrams
$(j,n)$ in \cite{esjetp,es}).

\section{Discussion of Results}

In this paper we have described derivation and properties of
a compact representation \eq{renormvert} for the off-shell one-loop
electron-photon vertex in the Yennie gauge, which is  convenient in
multiloop calculations. In practice of such calculations one usually
treats different terms in the integral representation of the
electron-photon vertex separately. Hence, it is not sufficient to have
a vertex with overall smooth asymptotic behavior but it is important to
have well behaved individual terms in the integral representation for
the vertex. We have specifically tailored these individual terms in
such way that they are described by the well behaved finite integrals
both in the infrared and ultraviolet regions. Respective integrals were
briefly discussed above, because control of their behavior is
absolutely crucial for successful applications when the one-loop vertex
plays the role of a subdiagram in multiloop diagrams (see, e.g.,
\cite{esjetp,es}).

One- and two-loop renormalization in the Yennie gauge was
considered earlier by G.Adkins \cite{adkins1,adkins2,adkins3,adkins4}.
In the framework of dimensional regularization he had obtained
interesting integral representations for the Yennie gauge one-loop
electron-photon vertex \cite{adkins1,adkins3}, which differ from the
representations considered above. One-loop electron-photon vertex in an
arbitrary gauge was also calculated in terms of elementary functions
and dilogarithms in \cite{kizi}. The vertex obtained in \cite{kizi} 
was written as a sum of longitudinal and transverse 
parts which were further decomposed into sums of different spinor 
structures. However, due to singular nature of separate analytic terms 
in \cite{kizi}, this nice representation is hardly suitable for the 
kind of applications we have in mind and partially discussed above.

The integral representations for the
Yennie gauge electron-photon vertex in \eq{renormvert} and
\eq{vertuvimp} were extensively used in our calculations of radiative
corrections of order $\alpha^2(Z\alpha)^5$ to hyperfine splitting and
Lamb shift \cite{esjetp,es}, and  in calculations of radiative-recoil
corrections of order $\alpha(Z\alpha)^5(m/M)$ to the Lamb shift
\cite{egs00}. Soft infrared behavior in the Yennie gauge greatly
facilitates the calculations. For example, almost all of the nineteen
diagrams with two radiative photon insertions in the electron line and
two external photon lines in \cite{esjetp,es} are infrared divergent in
the Feynman gauge. In the Feynman gauge calculations of the radiative
corrections of order $\alpha^2(Z\alpha)^5$ to hyperfine splitting and
Lamb shift induced by these diagrams are greatly impeded by the
infrared divergences though attainable \cite{kn1,kn10,pach2}. Due to
absence of the infrared divergences in the Yennie gauge and convenient
form of the vertex in \eq{renormvert} and \eq{vertuvimp} the results of
our calculations of the same contributions are about two orders of
magnitude more accurate than the results in
\cite{kn1,kn10,pach2,plwhkh}. We hope that the Yennie gauge off-shell
electron-photon vertex in \eq{renormvert} and \eq{vertuvimp} will find
further useful applications.

\acknowledgements

We are deeply grateful to H. Grotch for useful discussions and
suggestions.

This work was supported by the NSF grant PHY-0049059. Work of V. A.
Shelyuto was also supported in part by the RFBR grant \# 00-02-16718.

\end{document}